\RequirePackage[T1]{fontenc}
\documentclass[12pt]{article}
\usepackage[height=8.85in,width=6.45in]{geometry}
\usepackage[whole]{bxcjkjatype} 
\usepackage[utf8]{inputenc}
\usepackage{amsmath}
\usepackage{amssymb}
\usepackage{mathtools}
\numberwithin{equation}{section}
\usepackage{slashed}
\usepackage{braket}
\usepackage[svgnames]{xcolor}
\usepackage[colorlinks,linktocpage=true,citecolor=DarkGreen,linkcolor=FireBrick]{hyperref}
\usepackage{cite}
\usepackage{graphicx}
\usepackage{tikz}
\usepackage{tikz-cd}
\usepackage{times}
\usepackage[scaled]{couriers}
\usepackage{bm}
\usepackage{subfig}
\usepackage{caption}
\usepackage{mathrsfs}

\usepackage{xcolor}
\usepackage{mdframed}

\usepackage{ulem}

\renewenvironment{figure}[1][]{
  \begin{originalfigure}[#1]
    \begin{mdframed}[linecolor=black!0,backgroundcolor=black!0]%backgroundcolor=black!aについてa=0の時, 背景白色 a=1の時, わずかに灰色
}{
    \end{mdframed}
  \end{originalfigure}
}
\DeclareMathOperator{\Tr}{Tr}
\DeclareMathOperator{\tr}{tr}
\DeclareMathOperator{\Det}{Det}
\def\Re{\mathop{\mathrm{Re}}}

\def\diag{\mathop{\rm diag}\nolimits}

\usepackage{slashed}

\def\cA{{\cal A}}

\def\cC{{\cal C}}
\def\cD{{\cal D}}

\def\cF{{\cal F}}

\def\cL{{\cal L}}

\def\bR{{\mathbb R}}

\def\bZ{{\mathbb Z}}

\def\U{\mathrm{U}}
\def\SU{\mathrm{SU}}

\def\beq#1\eeq{\begin{align}#1\end{align}}

\def\i{{\mathsf i}}

\def\log{\text{log}}
\def\i{\mathsf i}
\def\calq{\mu}
\def\bosf{b}

\usepackage{comment} %new commandの\commentとconflictするのでこんなところに置いてます.
\newcounter{app}%Appendix用のcounter

\makeatletter
\@addtoreset{section}{app}%%%(*2)
\makeatother

\begin{document}

\begin{titlepage}

\begin{flushright}
TU-1188
\end{flushright}

\vskip 3cm

\begin{center}

{\Large \bfseries The QCD phase diagram in the space of imaginary chemical potential via 't~Hooft anomalies}

\vskip 1cm
Shun K. Kobayashi,  %\footnote{shun.kobayashi.p2@dc.tohoku.ac.jp}, 
Takahiro Yokokura,   %\footnote{takahiro.yokokura.t3@dc.tohoku.ac.jp}
and Kazuya Yonekura
\vskip 1cm

\begin{tabular}{ll}
Department of Physics, Tohoku University, Sendai 980-8578, Japan
\end{tabular}

\vskip 1cm

\end{center}

\noindent

The QCD phase diagram in the space of temperature and imaginary baryon chemical potential has been an interesting subject in numerical lattice QCD simulations because of the absence of the sign problem and its deep structure related to confinement/deconfinement. We study constraints on the phase diagram by using an 't~Hooft anomaly. The relevant anomaly is an anomaly in the space of imaginary chemical potential. We compute it in the UV, and discuss how it is matched by the pion effective field theory at low temperatures. Then we study implications of the anomaly to the phase diagram. There must be a line of phase transition studied in the past by Roberge and Weiss such that the expectation value of the Polyakov loop is not smooth when we cross the line. Moreover, if the greatest common divisor of the color and flavor numbers is greater than one, the phase transition across the Roberge-Weiss line must be either a first order phase transition, or a second order phase transition described by a nontrivial interacting three-dimensional CFT.

\end{titlepage}

\setcounter{tocdepth}{2}
%\tableofcontents

\newpage

\tableofcontents

\section{Introduction}

Confinement and chiral symmetry breaking are very important properties of Quantum Chromodynamics (QCD).
However it is difficult to analyze these phenomena based on the asymptotic freedom alone since they occur in the low energy, strongly coupled regime.
As a result, the phase transitions of confinement/deconfinement, chiral symmetry breaking, and their relation have been a difficult but interesting subject, and 
studied by various methods  (see e.g. \cite{Fukushima:2010bq,Bazavov:2011nk}).

Numerical lattice Monte Carlo simulations 
have produced many significant results.
However this method has  difficulty in some parameter regions, such as massless quark limits (i.e. chiral limits) and finite baryon chemical potential. 

Instead of numerical lattice simulations, we can also study the system by some analytical methods. One of the important methods is to use symmetries. When we study phase diagrams, symmetries play a key role as in the Landau's characterization of phases. We can distinguish phases by whether a symmetry is spontaneously broken or not.

Under some reasonable assumptions, 't Hooft showed that the chiral symmetry is spontaneously broken at zero temperature \cite{tHooft:1979rat}.
He used the method which nowadays is called 't~Hooft anomaly matching, and it has been applied to many strongly coupled systems.
't Hooft anomalies are the anomalies of global symmetries that would appear if the symmetries are gauged. The important and useful property of 't Hooft anomalies is the renormalization group (RG) invariance. The UV and IR theories must have the same 't~Hooft anomalies. Therefore the low energy effective action of a theory that has an 't Hooft anomaly in the UV must have some degrees of freedom that generates the 't Hooft anomaly. By this method, we can obtain rigorous results without solving strongly-coupled theories completely.

In recent years, 't Hooft anomalies have also been extended to the generalization of global symmetries~\cite{Gaiotto:2014kfa} and applied to various gauge theories,
including e.g.~\cite{Gaiotto:2017yup, Gaiotto:2017tne, Gomis:2017ixy,Komargodski:2017keh,Shimizu:2017asf,Komargodski:2017smk,Tanizaki:2017qhf,Tanizaki:2017mtm,Kikuchi:2018gfo,Tanizaki:2018wtg,Anber:2018iof,Yamaguchi:2018xse,Yonekura:2019vyz,Nishimura:2019umw,Wan:2019oyr,Bolognesi:2019fej,Wan:2019oax,Furusawa:2020qdz,Chen:2020syd,Bolognesi:2021yni,Tanizaki:2022ngt,Yamada:2022imq,Tanizaki:2022plm,Morikawa:2022liz,Bolognesi:2023sxe}. 
In addition, there is also another generalization of 't Hooft anomalies which involve parameter spaces of coupling constants, and they are used to study dynamics of strongly coupled systems~\cite{Gaiotto:2017yup,Kikuchi:2017pcp,Cordova:2019jnf,Cordova:2019uob}. This generalization is named anomalies in the space of coupling constants~\cite{Cordova:2019jnf,Cordova:2019uob}. 
This kind of anomalies implies that the system has some phase transition when we vary coupling constants.

In this paper, we investigate the phase diagram in the space of temperature and imaginary baryon chemical potential $(T,\mu_B)$. 
QCD in the presence of imaginary baryon chemical potential was studied by Roberge and Weiss \cite{Roberge:1986mm}.
Unlike real chemical potential, imaginary chemical potential does not have the sign problem, and hence numerical lattice Monte Carlo simulation is possible \cite{deForcrand:2002hgr}. For recent work, see e.g.~\cite{Bonati:2018fvg ,Goswami:2019exb, Guenther:2020jwe, Guenther:2022wcr, Cuteri:2022vwk ,DAmbrosio:2022kig} and references therein.\footnote{In particular, \cite{Bonati:2018fvg} summarizes many past works.}
One of the motivations to study the phase diagram involving imaginary chemical potential is that it contains theoretically interesting 
information related to confinement. In finite temperature QCD, it is difficult to define the concept of confinement/deconfinement rigorously due to the existence of dynamical quarks.
However, in the presence of imaginary chemical potential, we can get some information of confinement/deconfinement transition. There is a line of phase transition
on the $(T,\mu_B)$ plane across which the expectation value of the Polyakov loop is not smooth~\cite{Roberge:1986mm}. We may call this line as Roberge-Weiss line.
How Roberge-Weiss line behaves in the phase diagram contains information of confinement/deconfinement because the Polyakov loop is the order parameter for these phases. See \cite{Shimizu:2017asf,Yonekura:2019vyz} for more discussions.
We will use the anomaly matching method to get constraints on the phase diagram.

Let us summarize the rest of the paper. In Section \ref{uvanomaly}, we discuss an 't~Hooft anomaly of thermal QCD. This is an anomaly in the space of ``coupling constants'', where  imaginary chemical potential plays the role of a coupling constant. 
In Section \ref{lowanomaly},  
we study how the anomaly is matched in the low energy effective field theory of pions.
In Section \ref{imppd}, we consider applications of the anomaly to the QCD phase diagram. First we review the behavior of the theory in the high temperature region of the phase diagram. Then we give some examples of scenarios for the phase diagram allowed by the anomaly. 
In particular, we discuss the nature of the Roberge-Weiss line.

\section{The anomaly of the QCD Lagrangian}\label{uvanomaly}

We consider a four-dimensional QCD-like theory with color $N_c$ and massless flavor $N_f$.
In this paper we use differential form notations, so a gauge field $A_\mu$ is written as a 1-form $A=A_\mu dx^\mu$.
We use the convention that $A$ is anti-hermitian so that the covariant derivative is $D_\mu = \partial_\mu + A_\mu$.
The field strength of $A$ is $F = dA +A \wedge A$, or more explicitly $F = \frac12 F_{\mu\nu} dx^\mu \wedge dx^\nu$, where $F_{\mu\nu} = \partial_{\mu} A_{\nu} -\partial_{\nu} A_{\mu} + [A_\mu, A_\nu]$. We often omit the wedge product symbol $\wedge$ and write e.g. $F = dA+A^2$ and similarly for other differential form products.

The action is given by
\begin{align}
	S = -\frac{1}{2g^2}\int_{M_4}\tr_c(F_C\wedge \ast F_C) + \int_{M_4}d^4x\,\overline{\Psi}\gamma^\mu D_\mu \Psi,\label{eq:lagqcd}
\end{align}
where $\tr_c$ is the trace over gauge indices, $F_C$ is the curvature of the $\SU(N_c)$ gauge field $A_C$, $\ast$ is the Hodge star,\footnote{More explicitly $\int_{M_4}\tr_c(F_C\wedge \ast F_C)=\int_{M_4}d^4 x \frac{1}{2}\tr_c(F_{C\mu\nu}F_C^{\mu\nu})$.} $\Psi$ is the $N_f$ flavors of quarks which transforms in the fundamental representation of $\SU(N_c)$, $\gamma^\mu$ are the gamma matrices with $\{\gamma^\mu,\gamma^\nu\} = 2\delta^{\mu\nu}$, $g$ is the gauge coupling, and $M_4$ is the four dimensional (4d) spacetime which is always Euclidean in this paper.

Let $\SU(N_f)_L\times \SU(N_f)_R$ be the chiral symmetry that acts on the left handed quarks $\psi_L:=\frac{1+\overline{\gamma}}{2}\Psi$ and the right handed quarks $\overline{\psi}_R:=\frac{1-\overline{\gamma}}{2}\Psi$ as $\psi_L\mapsto g_L\psi_L$ and $\overline{\psi}_R\mapsto g_R\overline{\psi}_R$ for $g_{L/R}\in\SU(N_f)_{L/R}$, respectively.
Also let $\U(1)_V$ be the global symmetry under which $\Psi$ has charge 1. The combined gauge and global symmetry group of the action (\ref{eq:lagqcd}) is
\begin{align}
	 G=\frac{\SU(N_c)\times\SU(N_f)_L\times \SU(N_f)_R\times \U(1)_V}{\cC},\label{eq:symQCD}
\end{align}
where $\cC$ is the subgroup of $\SU(N_c)\times\SU(N_f)_L\times \SU(N_f)_R\times \U(1)_V$ consisting of elements which act trivially on the fields. 
The detail of $\cC$ will be discussed later.

The physical states of the QCD-like theory are color singlet (at least if the space is compact), meaning that $\SU(N_c)$ acts trivially on them.
Thus, in the context of physical states, the symmetry group (\ref{eq:symQCD}) is reduced to 
	\begin{align}
		H=\frac{\SU(N_f)_L\times \SU(N_f)_R\times \U(1)_B}{\cD},  \label{eq:symphy}
	\end{align}
where $\U(1)_B$ is the baryon number symmetry given by
	\begin{align}
		\U(1)_B:=\U(1)/\bZ_{N_c},
	\end{align}
in which $\bZ_{N_c}$ is the subgroup of $\U(1)_V$ whose generator is $e^{2\pi\i/N_c}\in\U(1)_V$, $\cD$ is the subgroup of $\SU(N_f)_L\times \SU(N_f)_R\times \U(1)_B$ consisting of elements which act trivially on gauge invariant states. The detail of $\cD$ also will be discussed later. 

The reason that $\U(1)_B=\U(1)/\bZ_{N_c}$ appears is as follows. The transformation $\Psi \to e^{2\pi \i /N_c}\Psi$ under $\bZ_{N_c}$ is equivalent to 
the gauge transformation by $e^{2\pi \i /N_c} \in \SU(N_c)$. Notice also that the gluon fields transform trivially under it.
Therefore, gauge invariant states are invariant under $e^{2\pi \i /N_c} \in \bZ_{N_c} \subset \U(1)_V$.
Consequently, we can divide $\U(1)_V$ by $\bZ_{N_c}$.

\subsection{The imaginary chemical potential and the $2\pi$ shift invariance}
We consider finite temperature QCD, and hence the spacetime is taken to be 
\beq
S^1 \times M_3,
\eeq
where $S^1$ is the Euclidean time direction whose circumference is the inverse temperature $\beta=1/T$, and $M_3$ is a space manifold
such as $\bR^3$ or $T^3$. We assume that the size of $M_3$ is large enough compared to any other scales such as the temperature and the dynamical scale of QCD.
In other words, we are interested in the large volume (thermodynamic) limit. 

We also introduce the imaginary baryon chemical potential $\mu_B$.
In the path integral, $\mu_B$ is introduced as a holonomy of the $\U(1)_B$ symmetry around $S^1$,
\begin{align}
	\mu_B = \i\int_{S^1} A_B ,\label{defmu}
\end{align}
where $A_B=(A_B)_\mu dx^\mu$ is the background gauge field of the $\U(1)_B$ symmetry. 
This $A_B$ is coupled to the $\U(1)_B$ current.
The coupling term of $A_B$ and $\Psi$ is 
\begin{align}\label{eq:QBterm}
	-S_{\text{int}}=-\int \overline{\Psi} \frac{1}{N_c} \slashed{A}_B \Psi
	= -\int_{S^1}A_B \int d^3x \frac{1}{N_c} \overline{\Psi} \gamma^0 \Psi
	=  \i\mu_B Q_B ,
\end{align}
where we have taken into account the fact that the quark field $\Psi$ has the $\U(1)_B$ charge $\frac{1}{N_c}$.
The partition function in the presence of $\mu_B$ is given by
\begin{align}
	Z(T,\mu_B) = \Tr e^{-\beta H+\i\mu_BQ_B},
\end{align}
where $H$ is the Hamiltonian and the trace is over the Hilbert space of the theory quantized on $M_3$.

Now, let us consider gauge transformations of $\U(1)_B$,
\begin{align}
	A_B &\to A_B - \i d\lambda,\\
	\Psi &\to e^{\i \lambda/N_c}\Psi.
\end{align}
There is a condition for $\lambda$ to be a gauge transformation such that the partition function remains invariant.
The gauge transformation, $A_B \to A_B -\i d\lambda$ acts on $\mu_B$ as
\begin{align}
	\mu_B 
		= \i\int_{S^1} A_B 
		\to \i\int_{S^1} (A_B  -\i d \lambda)
		= \mu_B + \int_{S^1} d \lambda.		 
\end{align}
The transformation of the partition function is
\begin{align*}
	Z(T,\mu_B)
		&= \Tr e^{-\beta H+\i\mu_BQ_B}\\
		&\to \Tr e^{-\beta H+\i\mu_BQ_B +\i \int_{S^1} d \lambda  Q_B}.
		\stepcounter{equation}\tag{\theequation} 
\end{align*}
This is invariant when $\lambda$ satisfies
\begin{align}
	\int_{S^1} d \lambda \in 2\pi \bZ, \label{eq:U(1)gaugcondition}
\end{align}
because we have $Q_B \in \bZ$ for gauge invariant states. This condition means that $e^{\i \lambda}$ (but not necessarily $e^{\i \lambda/N_c}$) is single-valued on $S^1$.

In summary, the partition function has the invariance under $\mu_B \to \mu_B+2\pi$,
\begin{align}
	Z(T,\mu_B+2\pi) = Z(T,\mu_B). \label{eq:muBinv}
\end{align}

\subsection{The mixed anomaly}
Now we see that the above $2\pi$ shift invariance has a mixed 't~Hooft anomaly with the chiral symmetry.
This anomaly will be derived from the perturbative anomaly in 4d.

Let $A_{L,R}$ be background gauge fields of $\SU(N_f)_L\times \SU(N_f)_R$.
We take them to be independent of the time coordinate and also take their time components to zero, so they are (the pullback of) gauge fields on the three-dimensional space $M_3$.
We denote the partition function as a functional of $A_{L,R}$ (as well as $T$ and $\mu_B$) by
\begin{align}
	Z(T,\mu_B,A_L,A_R).
\end{align}

The perturbative anomaly in 4d is described by the anomaly polynomial 6-form $I_6$. (For a review of the anomaly polynomial, see e.g. \cite{Weinberg:1996kr}.)
First notice that the covariant derivatives acting on the left and right handed quarks are given by
\begin{align}
	D_L = d + \cA_L,\quad 
	D_R = d + \cA_R,
\end{align}
where
\begin{align}
	\cA_L =A_C\otimes 1_{N_f}+ 1_{N_c}\otimes A_L + \frac{1}{N_c}1_{N_c}\otimes 1_{N_f}\cdot A_B,\label{eq:A_Ldef}\\
	\cA_R = A_C\otimes 1_{N_f}+ 1_{N_c}\otimes A_R + \frac{1}{N_c}1_{N_c}\otimes 1_{N_f}\cdot A_B.
\end{align}
The anomaly polynomial 6-form $I_6$ is given by
\begin{align}
	I_6 =\frac{1}{3!}\tr \left(\frac{\i\cF_L}{2\pi}\right)^3 - \frac{1}{3!}\tr \left(\frac{\i\cF_R}{2\pi}\right)^3,\label{eq:I6}
\end{align}
where $\cF_{L,R} = d\cA_{L,R}+\cA_{L,R}^2$, and the traces are taken in the representations of the left and right handed quarks, respectively. 
The meaning of the anomaly polynomial is as follows. We consider $I_5$ and $I_4^\lambda$ which appear in the anomaly descent equations
\begin{align}
	I_6 &= d I_5,	\label{eq:I6=dI5} \\
	\delta_\lambda I_5  &= d I^\lambda_4.
\end{align}
where $\delta_\lambda$ represents the gauge transformation by $\lambda$. Then the partition function $Z$
as a functional of the background fields transforms as
\beq
\delta_\lambda \log Z = -2\pi \i I_4^\lambda.
\eeq 
In this way, $I_6$ characterizes the perturbative anomaly.\footnote{
For more precise treatment of anomalies, it would be better to consider anomalies in terms of invertible field theories in one-higher dimensions. 
We regard an anomalous theory as living on the boundary of a bulk invertible field theory so that the total system has a well-defined, 
gauge invariant partition function by anomaly inflow. The total partition function is
      \begin{align}
        Z_\text{full}[A_B, A_L, A_R]:=Z_\text{bulk}[A_B, A_L, A_R]Z_\text{bdry}[A_B, A_L, A_R],
      \end{align}
where $Z_\text{bulk}[A_B, A_L, A_R]$ and $Z_\text{bdry}[A_B, A_L, A_R]$ are the bulk and boundary partition functions, respectively. Then the bulk invertible field theory partition function is given by $Z_\text{bulk}[A_B, A_L, A_R] = \exp( 2\pi \i \int I_5)$. 
}

Now, we would like to consider the anomaly related to $\U(1)_B$.
In particular, we focus on the following term,
\begin{align}
	I_6 \supset \frac{\i^3}{2(2\pi)^3}F_B(\tr_f F_L^2 - \tr_f F_R^2) 
		= \frac{\i F_B}{2\pi}(dI_3)
		=d\left[\frac{\i}{2\pi}A_B(dI_3)\right],
\end{align}
where
\begin{align}
	I_3 =  \frac{\i^2}{2(2\pi)^2} 
		\left[\tr_f \left(A_LdA_L+\frac{2}{3}A_L^3\right)
		- \tr_f \left(A_RdA_R+\frac{2}{3}A_R^3\right)
		\right].
\end{align}
Here $\tr_f$ is taken in the fundamental representation of $\SU(N_f)_{L,R}$.
The term proportional to $F_B (dI_3)$ represents a mixed anomaly between $\SU(N_f)_L\times \SU(N_f)_R$ and $\U(1)_B$.

In the anomaly descent equations, we focus on the $\U(1)_B$ gauge transformation $\delta_\lambda A_B = -\i d\lambda$
with the transformation parameter $\lambda =\frac{2\pi }{\beta}\tau $, where $\tau$ is the coordinate of the $S^1$.
This transformation satisfies the condition (\ref{eq:U(1)gaugcondition}), and it corresponds to $\mu_B\to\mu_B+2\pi$.
The $I_5$ contains the term
\begin{align}\label{eq:I5}
	I_5 &\supset \frac{\i}{2\pi}A_B(dI_3),
\end{align}
and hence
\begin{align}
	\delta_\lambda I_5 
		= \frac{\i}{2\pi}( -\i d\lambda)(dI_3)
		=d\left[\frac{d\tau}{\beta}I_3\right].
\end{align}
Thus we get $I^\lambda_4 = \frac{d\tau}{\beta}I_3$. Therefore, we get the anomaly under $\mu_B \to \mu_B+2\pi$ as
\begin{align*}
	Z(T,\mu_B+2\pi,A_L,A_R) 
		&= Z(T,\mu_B,A_L,A_R) \exp\left[-2\pi \i \int_{S^1\times M_3}I^\lambda_4\right]\\
		&=Z(T,\mu_B,A_L,A_R)\exp\left[-2\pi \i \int_{M_3}I_3\right]\\
		&=Z(T,\mu_B,A_L,A_R)\exp\left[\i (CS(A_L)-CS(A_R))\right] ,\label{eq:anomalyUV}
		\stepcounter{equation}\tag{\theequation} 
\end{align*}
where $CS(A)$ is the Chern-Simons invariant,
\begin{align}
	CS(A) 
		&:= \frac{1}{4\pi}\int_{M_3}\tr_f \left(AdA+\frac{2}{3}A^3\right).
\end{align}
We see that the invariance (\ref{eq:muBinv}) is broken by the introduction of background gauge fields $A_{L,R}$. This is the mixed 't~Hooft anomaly of our interest.\footnote{
The bulk invertible field theory for this anomaly is given by $Z_\text{bulk}[\mu_B, A_L, A_R]=\exp(\i \int \mu_B dI_3)$.
}

\subsection{Consequences of the anomaly}\label{sec:consequence}
We have found the mixed anomaly between $\SU(N_f)_L\times\SU(N_f)_R$ and the shift $\mu_B \to \mu_B+2\pi$. 
We can regard $\mu_B$ as a parameter, or a ``coupling constant''. 
Thus this is an anomaly in the space of coupling constants \cite{Cordova:2019jnf}.

The usual 't Hooft anomalies of global symmetries lead to constraints on the dynamics of the theory.
On the other hand, 't Hooft anomalies in the space of coupling constants give constraints on phase diagrams
when those coupling constants or parameters are varied~\cite{Kikuchi:2017pcp,Cordova:2019jnf}.

Following \cite{Cordova:2019jnf}, we argue that the existence of the anomaly (\ref{eq:anomalyUV}) implies either of the following possibilities:
\begin{enumerate}
\item There is at least one phase transition point when $\mu_B$ is varied from $0$ to $2\pi$.
\item There are some gapless degrees of freedom, such as the NG bosons associated to chiral symmetry breaking.
\end{enumerate}
Before discussing this point, let us make the following remark. The finite temperature is represented by the $S^1$, and we regard it as a compactification of the spacetime.
Then we can perform Kaluza-Klein (KK) decomposition to obtain a theory in three dimensions (3d). We can take the low energy limit of this 3d theory. A phase transition mentioned in the above statement can be described as a phase transition of this 3d theory. It can be either a first or second order phase transition.
In the case of a first order phase transition, the ground state of the 3d theory is changed in a discontinuous way. In the case of a second order phase transition, some gapless degrees of freedom with infinite correlation length appear.

To argue the existence of some phase transition or gapless degrees of freedom, suppose on the contrary that the system stays in a single gapped ground state for all $\mu_B\in\left[0, 2\pi\right]$. Because 't Hooft anomaly is RG invariant, we can take the low energy limit which is just the ground state of the 3d theory. 
By the assumption of the existence of the gap, there is no degrees of freedom in this low energy limit and hence
the partition function of the theory is just given by a local action of background fields.\footnote{
We emphasize that we are focusing on the 3d theory after the KK reduction on $S^1$. For instance, the partition function contains the usual thermal free energy of the original four-dimensional theory. However, from the point of view of the 3d theory, the free energy is just a cosmological constant term, so it is given by a local action.
} The local action may contain Chern-Simons terms of the $\SU(N_f)_L\times \SU(N_f)_R$ background gauge fields,
\beq
\log Z = -S_{\rm low} \supset \i k_L  CS(A_L) + \i k_R CS(A_R), \qquad k_L, k_R \in \bZ.
\eeq
As long as the assumption is true, the Chern-Simons levels $(k_L, k_R)$ at the ground state do not change as we vary $\mu_B$ because the Chern-Simons levels are discrete parameters and cannot change continuously. 
This conclusion is in contradiction with (\ref{eq:anomalyUV}). Therefore the assumption is not valid and there is at least one $\mu_B\in\left[0, 2\pi\right]$ at which there is a phase transition, or there is some gapless degrees of freedom such as NG bosons so that the partition function is not a local action of the background fields. 
This concludes our argument.
We will apply this result to the phase diagram of thermal QCD with imaginary chemical potential in Section \ref{imppd}.

Before closing this section, we will discuss candidates for degrees of freedom which match the anomaly. First let us consider a 3d fermion as a basic example of a phase transition implied by the anomaly (\ref{eq:anomalyUV}). 
Consider a massive fermion $\psi$ in the fundamental representation of $\SU(N_f)$ (which is either $\SU(N_f)_L$ or $\SU(N_f)_R$). Its action is given by
			\begin{align}
				S=\int_{M_3}d^3x\,\overline{\psi}\left(\sigma^i D_i+m\right)\psi,\label{eq:mfree fermion}
			\end{align}
where $\sigma^i$ are the Pauli matrices and $D_i$ is the covariant derivative. 
After integrating out the fermion when $m \neq 0$, we get the low energy effective action of the background $\SU(N_f)$ gauge field $A$ given by
			\begin{align}
				\log Z =-S_{\rm low} \supset \i \left(\frac{m}{2|m|} + k_0\right) CS(A),	\label{eq:massflowCS}
			\end{align}
where $k_0$ is an arbitrary (counter)term which is present before integrating out the fermion. 
When we change the fermion mass across $m=0$, the Chern-Simons level changes discontinuously. 
This is one of the possibilities consistent with the anomaly \eqref{eq:anomalyUV} if $m$ is an appropriate function of $\mu_B$.

If we consider $N_c=1$, i.e. free fermions without any gauge field in 4d, the above scenario really happens. In that case,
the mass $m$ in 3d is simply given by $m=\mu_B-2\pi (n+\frac12)$ where $n \in \bZ$ labels KK modes, and only one KK mode with $n=0$ becomes gapless when $\mu_B $ is varied from $0$ to $2\pi$ with the phase transition point $\mu_B=\pi$.\footnote{The phase transition here is not of Landau-Ginzburg type, but a type which happens when the ground state goes from one invertible phase to another. More explicitly, it is characterized by the change of the Chern-Simons level.}
However, the situation is different when $N_c>1$. Depending on the greatest common divisor of $(N_c,N_f)$, we will exclude the possibility of this kind of phase transition in QCD,
and argue that the phase transition must be either a first order transition or a second order transition described by a nontrivial 3d conformal field theory (CFT) rather than free fermions.

Finally let us also consider the possibility of a topological degree of freedom.
One may think that a topological degree of freedom is also a candidate for anomaly matching.
But we can see that it is not possible by an argument similar to Section~5 of \cite{Garcia-Etxebarria:2017crf}.
In order to see it, let us consider the behavior of the partition function of a topological quantum field theory (TQFT) $Z_\text{TQFT}[\mu_B, A_L, A_R]$ under $\mu_B \to \mu_B+2\pi$. We start with the variation of the partition function of such a theory under a small deformation of the imaginary chemical potential.
It is given by
	\begin{align}
		\frac{Z_\text{TQFT}[\mu_B+\delta\mu_B, A_L, A_R]}{Z_\text{TQFT}[\mu_B, A_L, A_R]}=\left\langle\exp\left(\i\int\,\delta\mu_B \,j\right)\right\rangle, \label{VTQFT}
	\end{align}
where $\delta\mu_B$ is a small variation of the imaginary chemical potential and $j$ is a 3-form current coupled to $\mu_B$. 
The right hand side of \eqref{VTQFT} is given by an integral of a local polynomial of $\delta \mu_B, \mu_B$ and $A_{L,R}$ by the following reason.
The $j$ is a local operator, and a theory which has a large mass gap like a TQFT has correlation functions that are exponentially small at long distances. In particular, in the limit of an infinite mass gap, correlation functions have only contact terms (i.e., delta functions and their derivatives).
The right hand side of \eqref{VTQFT} is given by correlation functions of local operators, and hence
the difference
	\begin{align}
		S[\mu_B, A_L, A_R]:=\log\,Z_\text{TQFT}[\mu_B, A_L, A_R]-\log\,Z_\text{TQFT}[0, A_L, A_R]
	\end{align}
takes the form of a local polynomial action. Throughout this paper, we use renormalization or counterterms for anomalies such that, in the triangle anomaly $\U(1)_B-\SU(N_f)_L-\SU(N_f)_R$, only the current for $\U(1)_B$ is not conserved by the anomaly and hence $\SU(N_f)_L \times \SU(N_f)_R$ is preserved.
Then $S[\mu_B, A_L, A_R]$ must be invariant under $\SU(N_f)_{L}\times\SU(N_f)_R$ gauge transformations. There is no candidate for a local polynomial of $\mu_B, A_L, A_R$ that is invariant under $\SU(N_f)_{L}\times\SU(N_f)_R$ gauge transformations and that can reproduce the anomaly \eqref{eq:anomalyUV}. A naive candidate is 
\beq
\int \frac{ \mu_B}{2(2\pi)^2} \left[\tr_f \left(A_L dA_L +\frac{2}{3} A_L^3\right) - \tr_f \left(A_R dA_R +\frac{2}{3} A_R^3\right)\right],
\eeq
but this is not invariant under $\SU(N_f)_{L}\times\SU(N_f)_R$ gauge transformations;
see the discussions below \eqref{eq:anom matching term} for more comments. 
This means that a TQFT does not match the anomaly \eqref{eq:anomalyUV} and hence we exclude it as a possible candidate.

\section{The anomaly of the chiral Lagrangian}\label{lowanomaly}

		In this section, we will study how the anomaly is matched in the low energy effective theory of the QCD-like theory. In sufficiently low energy, the chiral symmetry $\SU(N_f)_L\times \SU(N_f)_R$ is spontaneously broken into its diagonal subgroup $\SU(N_f)_D$. The Nambu-Goldstone boson associated to the chiral symmetry breaking is the pion denoted by 
		\beq
		U(x)\in [\SU(N_f)_L\times\SU(N_f)_R ]/ \SU(N_f)_D \simeq \SU(N_f).
		\eeq
		The kinetic term of the effective action in four-dimensional Euclidean space $M_4=S^1\times M_3$ is 
			\begin{align}\label{eq:4dpionaction}
				S_{\text{4d}} \supset \int_{M_4} d^4x \, \frac{1}{2}f_\pi^2 \tr_f (D^\mu U^\dagger D_\mu U),
			\end{align}
		where $f_\pi$ is the pion decay constant, and $D_\mu$ is the covariant derivative 
			\begin{align}
				D_\mu U:=\partial_\mu U+A_{L\mu}U-UA_{R\mu}.
			\end{align}
We can expand the pion field into KK modes:
			\begin{align}
				U(x)=\sum_{n\in\mathbb{Z}}e^{2\pi \i n\tau/\beta}U_n(\vec{x}).
			\end{align}
As far as the anomaly \eqref{eq:anomalyUV} is concerned, we can focus on the low energy limit of the 3d theory and hence we take the KK modes with $n \neq 0$ to be zero. Thus the action becomes
			\begin{align}
				S_{\text{3d}}\supset \int_{M_3} d^3x \, \frac{\beta}{2}f_\pi^2 \tr_f (D^i U^\dagger D_i U).
			\end{align}
		where now $U(\vec{x})$ depends only on $M_3$. As mentioned earlier, $A_{L,R}$ are taken to be (the pullback of) gauge fields on $M_3$.

	\subsection{The anomaly matching term in the chiral Lagrangian }
		According to the 't Hooft anomaly matching condition, the low energy effective action of the QCD-like theory should reproduce the same anomaly as (\refeq{eq:anomalyUV}). The anomaly in 4d is known to be matched by the Wess-Zumino-Witten (WZW) term, and hence we expect that the reduction of the WZW term to 3d gives the necessary term for the anomaly matching. However, in the following, we will directly write down the relevant term.
		
		To treat $\SU(N_f)_L$ and $\SU(N_f)_R$ symmetrically, it is convenient to describe the pion by the coset construction.		
		We introduce two matrices $(U_L,U_R)\in\SU(N_f)_L \times\SU(N_f)_R$ on which the symmetry groups $\SU(N_f)_L$ and $\SU(N_f)_R$ act from the left, respectively. 
		To realize the coset, we also introduce a gauge symmetry $\SU(N_f)_H$, called the hidden local symmetry, which acts on both $U_L$ and $U_R$ from the right. 
		More explicitly, 	$g_L\in \SU(N_f)_L,g_R\in \SU(N_f)_R$, and $g_H\in \SU(N_f)_L$ act as
			\begin{align}
				(U_L, U_R)  \to (g^{-1}_L U_L g_H, g^{-1}_RU_R g_H)
			\end{align}
		In this way we can realize the coset $ [\SU(N_f)\times\SU(N_f)_R ]/ \SU(N_f)_D$.
The matrix that is invariant under the hidden local symmetry is given by
			\begin{align}
				U = U_LU_R^\dagger.
			\end{align}
This is the usual pion field which appears in \eqref{eq:4dpionaction}.			

		A slight advantage of using the hidden local symmetry is as follows. Let us introduce
			\begin{align}
				A_L^{U_L} &:=  U_L^{-1}A_LU_L+U_L^{-1}dU_L, \label{eq:A_L^{U^L}}\\
				A_R^{U_R} &:=  U_R^{-1}A_RU_R+U_R^{-1}dU_R. \label{eq:A_R^{U^R}}
			\end{align}
		$A_L$ and $A_R$ transform under $(g_L, g_R, g_H)$ as
			\begin{align}
				A_L &\to g_L^{-1}A_Lg_L +g_L^{-1}dg_L,\\
				A_R &\to g_R^{-1}A_Rg_R +g_R^{-1}dg_R.
			\end{align}
		Then $A_L^{U_L} $ and $A_R^{U_R}$ transform under $(g_L, g_R, g_H)$ as	
			\begin{align}
				A_{L,R}^{U_{L,R}} \to g_H^{-1}A_{L,R}^{U_{L,R}}g_H+g_H^{-1}dg_H. \label{eq:HLT}
			\end{align}
		We can see that both $A_{L}^{U_{L}}$ and $A_{R}^{U_{R}}$ transform in the same way. Therefore both $A_{L}^{U_{L}}$ and $A_{R}^{U_{R}}$ are connections of the same principal bundle whose structure group is $\SU(N_f)_H$. It is possible to fix the gauge by $U_L \to U$ and $U_R \to 1$, but then the treatment of $\SU(N_f)_L$ and $\SU(N_f)_R$ is a bit asymmetric.\footnote{The convenience of the hidden local symmetry is clearer if we consider more general theories. See \cite{Yonekura:2020upo} for very general situations.}

		Let us construct the anomaly matching term. For this purpose, we will define a 3-form $J_3$ such that
		\beq
		dJ_3 =  \frac{1}{2(2\pi)^2} 
		\left[\tr_f \left(F_L^2\right)
		- \tr_f \left(F_R^2\right)
		\right].
		\eeq
		By using $J_3$, we define the anomaly matching term by
			\begin{align}
				-S_{\text{anom}} = \int_{M_3}\i\,\mu_BJ_3. \label{eq:anom matching term}
			\end{align}
		If we take $J_3$ to be 	
		\beq
		 J_3 \overset{?}{\sim} \frac{1}{2(2\pi)^2} \left[\tr_f \left(A_L dA_L +\frac{2}{3} A_L^3\right) - \tr_f \left(A_R dA_R +\frac{2}{3} A_R^3\right)\right],
		 \eeq
		 then naively one might think that we can reproduce the anomaly  \eqref{eq:anomalyUV} under $\mu_B \to \mu_B +2\pi$.
		  If this were the case, the existence of some massless degrees of freedom like the pion would not have been necessary.
		  However, this is not correct. If we take $J_3$ as above, then \eqref{eq:anom matching term} would be a Chern-Simons term whose Chern-Simons level $\mu_B/2\pi$ is not quantized for generic values of $\mu_B$ and hence it is not gauge invariant under $\SU(N_f)_L \times \SU(N_f)_R$. 
		  		  We need to modify $J_3$ so that it is gauge invariant.\footnote{ 
				  We remark that, as far as we are concerned with only the triangle anomaly $\U(1)_B- \SU(N_f)_{L,R}-\SU(N_f)_{L,R}$, we can take a counterterm so that the $\SU(N_f)_{L,R}$ is gauge invariant. Then only the $\U(1)_B$ is anomalous. This is the choice we are using throughout this paper. This choice was implicit when we have taken $I_5$ to be as in \eqref{eq:I5} whose right hand side is invariant under gauge transformations of $\SU(N_f)_L \times \SU(N_f)_R$  but not of $\U(1)_B$.}				
		For this purpose, we need the pion field. 
		
		A definition of $J_3$ which will work is as follows. 
		Let us take a one parameter family of gauge fields given by
			\begin{align}
				A_t&:=(1-t)A_R^{U_R}+tA_L^{U_L},\label{eq:one para fam}
			\end{align}
		where $t$ is the parameter.	
		Then we define $J_3$ by
			\begin{align}
				J_3&=\frac{1}{(2\pi)^2}\int_0^1\,dt\,\tr_f(F_t\partial_tA_t),
			\end{align}
		where $F_t=dA_t+A_t^2$ is the curvature of the gauge field $A_t$.
		One can check that this is gauge invariant by using \eqref{eq:HLT}: 
			\begin{align*}
				 \tr_f(F_t\partial_tA_t)
				\rightarrow \tr_f\left[g_H^{-1}F_tg_H\partial_t\left(g_H^{-1}A_tg_H+g_H^{-1}dg_H\right)\right] =
				 \tr_f(F_t\partial_tA_t).
				\stepcounter{equation}\tag{\theequation} 
			\end{align*}	
		By a straightforward computation one can also check that
		\beq
		2\pi \int_{M_3} J_3 = CS\left(A_L^{U_L}\right)-CS\left(A_R^{U_R}\right)+\frac{1}{4\pi}\int_{M_3}d\,\tr_f\left(A_R^{U_R}A_L^{U_L}\right).
		\eeq
		The last term is zero since it is a total derivative.
		We also have
		\beq
		CS(A_{L,R}^{U_{L,R}}) =CS\left(A_{L,R}\right) \mod 2\pi \bZ.
		\eeq
		The reason is that we can regard $A_{L,R}^{U_{L,R}}$ as the gauge transformation of $A_{L,R}$ by $U_{L,R}$, and the Chern-Simons invariant is well-known to be gauge invariant modulo $2\pi \bZ$. Therefore, we get
		\beq
		2\pi \int_{M_3} J_3  = CS\left(A_L \right)-CS\left(A_R \right) \mod 2\pi \bZ.
		\eeq
		By using it, we see that (\ref{eq:anom matching term}) reproduces the anomaly \eqref{eq:anomalyUV} under $\mu_B \to \mu_B+2\pi$.
		
		There are other possible terms of the 3d effective action that are allowed by symmetry. However, they are not important for the anomaly matching, and for simplicity we do not explicitly write them. Then the 3d effective action is given by
			\begin{align}
				-S_{\text{3d}}=-\int_{M_3} d^3x \, \frac{\beta}{2}f_\pi^2 \tr (D^i U^\dagger D_i U)+\int_{M_3}\i\,\mu_B J_3. \label{eq:lageff}
			\end{align}
			As a check, let us consider the case that the background fields $A_{L,R}$ are set to zero. We can fix the gauge of $\SU(N_f)_H$ by setting $U_R \to 1$ and  $U_L \to U$. Then we get
			\beq
			\int_{M_3}\i\,\mu_B J_3 \to \int_{M_3}\i\,\mu_B \frac{-1}{24\pi^2} \tr_f (U^{-1}dU)^3 = \i\,\mu_B Q_B,
			\eeq
where we used the identification of the baryon charge $Q_B$ with $-\frac{1}{24\pi^2} \int \tr_f (U^{-1}dU)^3$~\cite{Witten:1983tx}. 	
The expression $\i\,\mu_B Q_B$ is as expected from \eqref{eq:QBterm}.

\section{Implications for QCD phase transition}\label{imppd}
In the previous sections we have studied the anomaly given by \eqref{eq:anomalyUV} and some of its consequences.
Now we consider its application to the QCD phase diagram on the plane $(T,\mu_B)$.

\subsection{The phase diagram at high temperatures}\label{sec:freeE}

We define the thermal free energy by
\begin{align}\label{eq:freeE}
	F(T,\mu_B) = -\frac{1}{\beta V_3}\log Z(T,\mu_B),
\end{align}
where $V_3 \to \infty$ is the volume of the 3d space $M_3$.
At high temperatures, we can calculate the free energy at the one-loop approximation~\cite{Gross:1980br}.
To give the results, let us first define the following function,
\begin{align}
	f(T,\mu) 
		&= \pi^2 T^4\left(-\frac{7}{8}\cdot \frac{1}{45} +\frac{1}{12}\left(\frac{\mu}{\pi}\right)^2 -\frac{1}{24}\left(\frac{\mu}{\pi}\right)^4\right)
		\quad \text{for}\ -\pi\leq \mu \leq \pi, \label{eq:ferfdef}\\
	f(T,\mu) 
		&= f(T,\mu+2\pi) .
\end{align}
We remark that this function is not smooth at $\mu=\pi$.

The free energy of a massless Weyl fermion $\psi$ (with the anti-periodic boundary condition $\psi(\beta)=-\psi(0)$ on $S^1$) which is coupled to the chemical potential $\mu$ is given by 
$f(T,\mu)$. On the other hand, the free energy of a massless complex boson $\phi$ (with the periodic boundary condition $\phi(\beta) = \phi(0)$ on $S^1$) is given by $\bosf(T,\mu)$, which is defined as
\begin{align}
	\bosf(T,\mu) 
		&= -f(T,\mu-\pi)\notag\\
		&= -\pi^2 T^4 \left(\frac{1}{45}-\frac{1}{6}\left(\frac{\mu}{\pi}\right)^2+\frac{1}{6}\left(\frac{\mu}{\pi}\right)^3-\frac{1}{24}\left(\frac{\mu}{\pi}\right)^4\right)
			\quad \text{for}\ 0\leq \mu \leq 2\pi
			\label{eq:bosfdef}\\
			\bosf(T,\mu) &= \bosf(T,\mu+2\pi) 
\end{align}
We give the details of their calculations in Appendix \ref{sec:freeenergy}.

In our case of the QCD-like theory,
the gauge field which is coupled to the quarks has the following form,
\begin{align}\label{eq:C+B}
	A_C+\frac{1}{N_c}1_{N_c}A_B.
\end{align}
The Wilson line of this gauge field around $S^1$, up to gauge transformations, is given by
\begin{align}
	{\rm P}\exp\left( - \int_{S^1} \left(A_C+\frac{1}{N_c}1_{N_c}A_B\right)  \right) = \diag(e^{\i a_1},\cdots, e^{\i a_{N_c}}), \label{eq:quarkWilsonline}
\end{align}
where $a_i$ satisfy
\begin{align}
	\sum_{i=1}^{N_c} a_i \in \mu_B +2\pi \bZ .	\label{eq:aiconstraint}
\end{align}
To understand this constraint, recall the definition of $\mu_B$ given in (\ref{defmu}).
We get (\ref{eq:aiconstraint}) by taking the determinant of each side of (\ref{eq:quarkWilsonline}) and use the fact that $A_C$ is the gauge field of $\SU(N_c)$.

We first consider the free energy as a function of constant $\{a_i\}$ as in the computation of Coleman-Weinberg potentials,
\begin{align}
	F(T,\{a_i\}) = F_{\text{gluon}}(T,\{a_i\})+F_{\text{quark}}(T,\{a_i\})	\label{eq:TotalF}
\end{align}
where $F_{\text{gluon}}(T,\{a_i\})$ and $F_{\text{quark}}(T,\{a_i\})$ are the one-loop contributions of the gluons and quarks around the point $\{a_i\}$.
Then, we obtain $F(T,\mu_B)$ by minimizing $F(T,\{a_i\}) $ with respect to $\{a_i\}$ under the condition \eqref{eq:aiconstraint}.

The gluons and quarks are in the adjoint and fundamental representations of $\SU(N_c)$, respectively.
Then, by using \eqref{eq:ferfdef} and \eqref{eq:bosfdef}, $F_{\text{gluon}}(T,\{a_i\})$ and $F_{\text{quark}}(T,\{a_i\})$ are given by
\begin{align}
	F_{\text{gluon}}(T,\{a_i\}) 
		&= (N_c-1)\bosf(T,0) + \sum_{i\neq j}\bosf(T,a_i-a_j)\label{eq:gluonF}\\
	F_{\text{quark}}(T,\{a_i\})
		&=2N_f\sum_i f(T,a_i)\label{eq:quarkF}
\end{align}
See Appendix \ref{sec:freeenergy} for the details. We remark that a single gauge field $A_\mu$ contains two particles (i.e. two helicity modes)
which is the same number of particles in a single complex boson.

\begin{figure}
    \begin{tabular}{cc}
      \begin{minipage}[t]{0.45\hsize}
        \centering
        \includegraphics[keepaspectratio, scale=0.45]{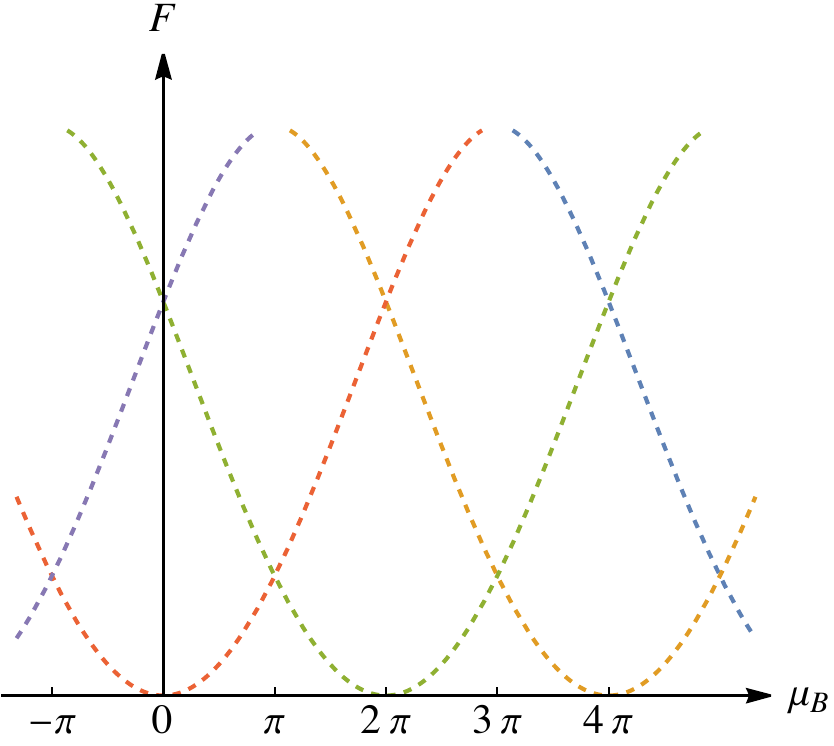}
        \caption{The free energy at (\ref{eq:a=a}) for several values of $n$. For example, the red dashed line is $n=0$, the green dashed line is $n=-1$.}
        \label{fig:freeenergy}
      \end{minipage} &
      \begin{minipage}[t]{0.45\hsize}
        \centering
        \includegraphics[keepaspectratio, scale=0.45]{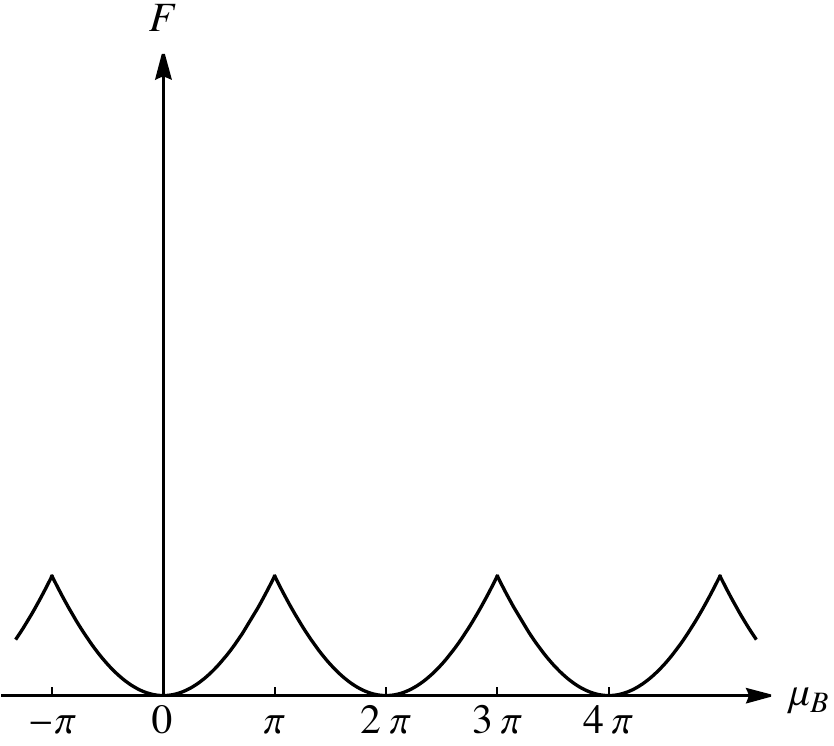}
        \caption{The true free energy. It has cusps at $\mu_B = (2m+1)\pi, m \in \bZ$.}
        \label{fig:truefreeenergy}
      \end{minipage}
    \end{tabular}
  \end{figure}

The values of $\{a_i\}$ at which $F(T,\{a_i\}) $ is minimized are
\begin{align}
	a_1 = \cdots = a_{N_c} = \frac{\mu_B+2\pi n}{N_c},   	\label{eq:a=a}
\end{align}
where $n \in \bZ$ is an integer such that $- \pi \leq \mu_B+2\pi n \leq \pi$. See Appendix~\ref{sec:min} for a proof.
Within the range $-\pi<\mu_B<\pi$, the minimum value is achieved at $n=0$.
When $\mu_B=\pi$, the values of $F_{\text{gluon}}(T,\{a_i\}) $ at $n=0$ and $n=-1$ are the same.
In Figure \ref{fig:freeenergy}, we show a graph of the free energy for several values of $n$. In Figure \ref{fig:truefreeenergy}, we show the true free energy, which is the minimum of Figure  \ref{fig:freeenergy}. The free energy has cusps at $\mu_B = (2m+1)\pi, m \in \bZ$.
This implies that a first order phase transition occurs from one vacuum (corresponding to $n=-m$) to another (corresponding to $n=-m-1$) when $\mu_B$ crosses $(2m+1)\pi$ at high temperatures.

At high temperatures, there are no gapless degrees of freedom in the 3d theory by the following reason.
The quarks have the anti-periodic boundary condition and their KK modes are all massive.
The time components of the gluons $(A_C)_4$ are massive because they are described by $\{a_i\}$ and have the potential energy
given by $F(T,\{a_i\}) $. It is interpreted as the Coleman-Weinberg potential for $\{a_i\}$. Finally, the components of the gluons
in the 3d space direction becomes a 3d $\SU(N_c)$ Yang-Mills theory which is believed to be gapped with a unique vacuum. Therefore, all the fields are gapped at high temperatures. 
To get gapless degrees of freedom at low temperatures, we need some phase transition. For example, we should have a chiral phase transition so that the chiral symmetry is broken at low temperatures. 

\subsection{Constraints on the phase diagram}
		\begin{figure}
			\centering
			\includegraphics[scale=0.4]{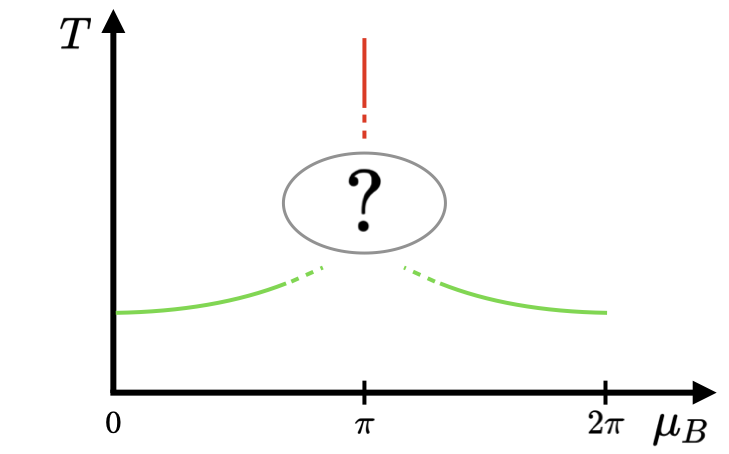}
			\caption{A summary of sufficiently high or low temperature regions. The green line represents chiral phase transition below which the chiral symmetry is broken. The red line is a phase transition 
			which is detected by the Polyakov loop operator \eqref{eq:polyakov}. }
			\label{fig:pd?}
		\end{figure}
	We summarize the phase diagram in the sufficiently high and low temperature regions in Figure~\ref{fig:pd?}. The red line of Figure~\ref{fig:pd?} is the phase boundary which shows the first-order phase transition as described in Subsection~\ref{sec:freeE}. In this first order phase transition,
the vacuum expectation value $\braket{L} $ of the Polyakov loop operator $L$ of the $\SU(N_c) \times\U(1)_B$ gauge field~\footnote{More precisely, the group is $[\SU(N_c) \times \U(1)_V]/\bZ_{N_c}$.} jumps discontinuously. The Polyakov loop operator $L$ is given by
\beq\label{eq:polyakov}
L =\tr_c 	{\rm P}\exp\left( - \int_{S^1} \left(A_C+\frac{1}{N_c}1_{N_c}A_B\right)  \right) .
\eeq
At the level of the one-loop computations of the previous section, we have
\beq
\braket{L} = \sum_{i=1}^{N_c}e^{\i a_i} = N_c \exp \i \left(  \frac{\mu_B+2\pi n}{N_c} \right),
\eeq
where we have used \eqref{eq:a=a}.
 The green line of Figure~\ref{fig:pd?} is a phase boundary which shows the chiral phase transition~\footnote{We only consider $(N_c, N_f)$ such that the chiral symmetry is broken at sufficiently low temperatures.}. It is difficult to determine what happens at intermediate temperatures analytically because this region requires the understanding of strongly coupled nonperturbative effects. Instead of solving QCD completely or numerically, we will rule out some logical possibilities and discuss some examples that are not ruled out.
 
\begin{figure}
			\centering
			\includegraphics[scale=0.4]{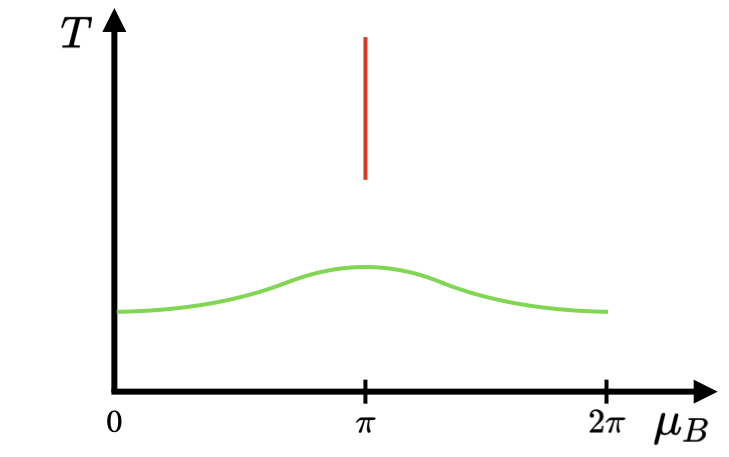}
			\caption{A typcial example which is excluded by the anomaly. The green line represents chiral phase transition below which the chiral symmetry is broken. The red line is a phase transition 
			which is detected by the Polyakov loop operator \eqref{eq:polyakov}. }
			\label{fig:pdimp}
		\end{figure}

\begin{figure}[tbp]
	\subfloat[The case which remains first order.]{\label{fig:pdpt}\includegraphics[scale=0.32]{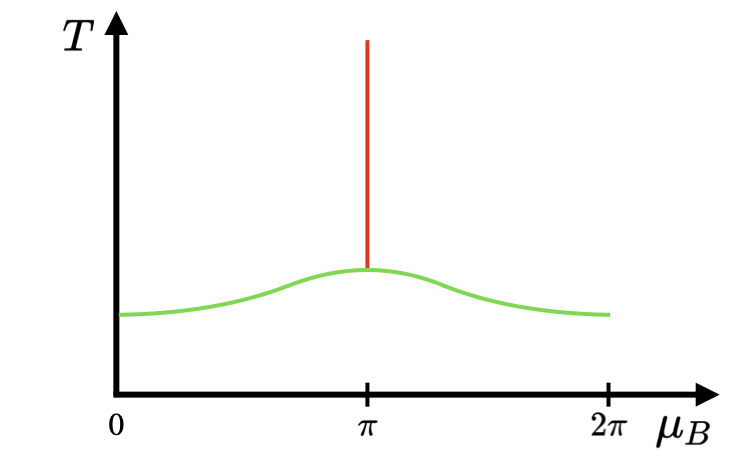}}
	\subfloat[The case which changes to second order.]{\label{fig:pdps}\includegraphics[scale=0.32]{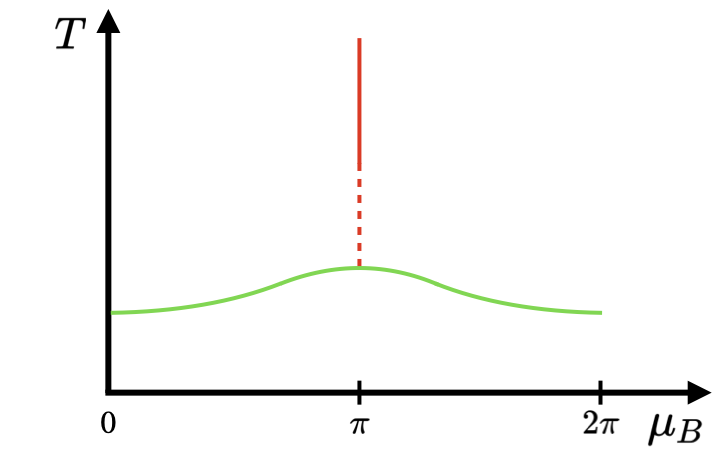}}
	\caption{Two typical examples for the middle region. In the left figure, the phase transition on the red line remains first order until it reaches the green line representing chiral phase transition. In the right figure, the phase transition is changed from first order to second order. The red dashed line represent a second order phase transition. If $\text{gcd}(N_c, N_f)>1$, the second order transition cannot be described by 3d free fermions. We need some nontrivial CFT. 
	}
	\label{fig:pospd}
\end{figure}

There are many logical possibilities for the phase diagram. However, we just focus on typical diagrams shown in Figure~\ref{fig:pdimp} and \ref{fig:pospd}. In these diagrams, we take into account only the minimum number of phase boundaries and do not introduce anything which is not required by symmetries and anomalies. However, the argument is similar for more complicated cases. 

	First, by the result in Section~\ref{uvanomaly}, we can rule out the cases which have a region without any phase transition nor chiral symmetry breaking as in Figure~\ref{fig:pdimp}. The reason is as follows. At high temperatures, the 3d theory is gapped as discussed in the previous section. It remains gapped as long as there are no phase transitions. (Appearance of new gapless degrees of freedom is, by definition, a phase transition.) In Figure~\ref{fig:pdimp}, we can take a path from $(T_0, \mu_B=0)$ to $(T_0, \mu_B=\pi)$ for some large enough $T_0$ such that we encounter no phase transition along the path.
This is in contradiction with the existence of the anomaly \eqref{eq:anomalyUV} as discussed in Subsection~\ref{sec:consequence}.

	Second, we will discuss constraints on the nature of the phase transition at the red line of Figure~\ref{fig:pospd}.
Two examples are shown in the Figure~\ref{fig:pospd}. 
As mentioned before, there is a first order phase transition on the red line if the temperature is sufficiently high. 
In the case of Figure~\ref{fig:pdpt}, the phase transition remains first order
until it reaches the green line of chiral symmetry breaking. In the case of Figure~\ref{fig:pdps}, the phase transition is changed from first to second order at some point on the red line. The dashed part of the red line represents a second order phase transition. 

One of the possibilities for the second order phase transition on the red dashed line of Figure~\ref{fig:pdps} is that some 3d free fermions become gapless on the red dashed line. Then we can match the anomaly as discussed in Subsection~\ref{sec:consequence}. This case is realized in some examples of gauge theories.\footnote{
For instance, this case is realized in $N_f=N_c+1$ supersymmetric QCD~\cite{Intriligator:1995au}. In that case, the free 3d fermions come from composite baryon particles in 4d. 
} 
However, if the greatest common divisor $n:=\text{gcd}(N_c, N_f)$ is greater than unity, $n>1$, then we can exclude the possibility of 3d free fermions
as we are now going to discuss.

To exclude 3d free fermions,
	we will use the global structure of the symmetry of QCD given in (\ref{eq:symQCD}) and \eqref{eq:symphy}. Notice that there is the subgroup $\cC$ in \eqref{eq:symQCD} which acts trivially on any field of the theory. The generators of $\cC$ are given by 
		\begin{align}
			&c_1=(e^{2\pi\i/N_c}, 1, 1, e^{-2\pi\i/N_c})\in\SU(N_c)\times\SU(N_f)_L\times \SU(N_f)_R\times \U(1)_V,\\
			&c_2=(1, e^{2\pi\i/N_f}, e^{2\pi\i/N_f}, e^{-2\pi\i/N_f})\in\SU(N_c)\times\SU(N_f)_L\times \SU(N_f)_R\times \U(1)_V.
		\end{align}
The global symmetry group (\ref{eq:symphy}) that acts on gauge invariant states and operators is obtained by removing $\SU(N_c)$.
The group $\cD$ which appears in that equation is generated by 
		\begin{align}
			c_2^\prime:=(e^{2\pi\i/N_f}, e^{2\pi\i/N_f}, e^{-2\pi\i N_c/N_f})\in\SU(N_f)_L\times \SU(N_f)_R\times \U(1)_B,
		\end{align}
where $\U(1)_B=\U(1)_V/\bZ_{N_c}$.

Now the point is as follows. 3d free fermions are, by definition, neutral under gauge symmetries.\footnote{We will briefly discuss the case that 3d fermions are charged under some gauge symmetries.}
Then they must be in some representation of the global symmetry group \eqref{eq:symphy} and in particular they are invariant under $\cD$. We will show that such representations which are neutral under $\cD$ cannot reproduce the anomaly \eqref{eq:anomalyUV} if $n=\text{gcd}(N_c, N_f)>1$ . Recall that the contribution of a 3d fermion in a representation $\rho$ to the anomaly is given by 
\beq
\pm \frac{\i}{4\pi} \tr_\rho \left(AdA+\frac{2}{3}A^3 \right),
\eeq
where the trace $\tr_\rho$ is taken in the representation $\rho$, and the sign depends on whether the 3d fermion mass crosses zero from positive to negative values or the other way around.
See \eqref{eq:massflowCS} which is written for the case of the fundamental representation.

Any reducible representation is given by a sum of irreducible representations, so we consider irreducible representations. Also,
we are not interested in the baryon charges of 3d fermions, so we focus on the subgroup
\beq\label{eq:Zn}
\frac{\SU(N_f)_L\times\SU(N_f)_R}{\bZ_n} \subset \frac{\SU(N_f)_L\times\SU(N_f)_R \times \U(1)_B}{\cD},
\eeq
where the $\bZ_n$ on the left hand side is generated by
\beq
c_3 :=(c'_2)^{N_f/n}= (e^{2\pi\i/n}, e^{2\pi\i/n}, 1)\in\SU(N_f)_L\times \SU(N_f)_R\times \U(1)_B.
\eeq
In other words, we consider representations $\rho_L\otimes\rho_R$ of $\SU(N_f)_L\times \SU(N_f)_R$ which are invariant under the above $\bZ_n$.

For instance, the simplest case in which the Chern-Simons level is changed by 1 as required by the anomaly \eqref{eq:anomalyUV} is the fundamental representation $\rho_f$ of $\SU(N_f)_L$ or $\SU(N_f)_R$. More precisely, we consider $\rho_f \otimes 1$ or $1 \otimes \rho_f$, where $\rho_f$ is the fundamental representation and $1$ is the trivial representation. However, this possibility is excluded since the fundamental representation is not invariant under $\bZ_n$ if $n>1$. 
In the following, we argue that no representation of $\SU(N_f)_L\times \SU(N_f)_R/\bZ_n$ can change the Chern-Simons level by 1.

Given a representation $\rho_L$ of $\SU(N_f)_L$,
let $k_L $  be the Dynkin index of $\rho_L$ normalized so that $k_L=1$ for the fundamental representation $\rho_f$. Explicitly we have $k_L := \tr_{\rho_L} T_L^2/\tr_{\rho_f} T_L^2 $, where $T_L$ is any Lie algebra generator.
Also let $d_L$ be the dimension of $\rho_L$. We also define $k_R$ and $d_R$ in the same way for a representation $\rho_R$ of $\SU(N_f)_R$.
The change of the Chern-Simons levels of $A_L, A_R$ when the mass of a 3d fermion in an irreducible representation $\rho_L\otimes\rho_R$ is varied from negative to positive values is
\begin{align}
	k_L d_R CS(A_L) + k_R d_L CS(A_R).
\end{align}
This is because, for any Lie algebra generator $T_L$ of $\SU(N_f)_L$, we have
\begin{align}
	\frac{\tr_{\rho_L\otimes \rho_R} T_L^2}{\tr_{\rho_f\otimes 1} T_L^2}  = d_R \frac{\tr_{\rho_L} T_L^2}{\tr_f T_L^2} = d_Rk_L.
\end{align}

In Appendix~\ref{sec:technicalproof}, we prove that $k_Ld_R $ is an integer multiple of $ n$, 
\beq\label{eq:multn}
k_Ld_R \in n\bZ.
\eeq
Then the change of the Chern-Simons level is always a multiple of $n$. This is true even if 3d fermions are in some reducible representations because in that case the change of the Chern-Simons level is given by the sum of contributions of irreducible representations. 
Therefore, the anomaly \eqref{eq:anomalyUV} cannot be matched by using free 3d fermions when $n>1$.

It is logically possible that the second order phase transition at the red dashed line of Figure~\ref{fig:pdps} is described by a 3d gauge theory in which fermions are charged under some gauge group.
If 3d fermions are charged under a continuous gauge group, the second order phase transition is described by a nontrivial 3d CFT involving that gauge interaction.
Another possibility is that the gauge group of the low energy effective theory is a discrete group in which case the gauge degrees of freedom are topological field theory. However, in the specific case of Figure~\ref{fig:pdps}, the possibility of topological degrees of freedom may be excluded as follows. If we are slightly away from the red line, then there are no such topological degrees of freedom by the following reason. The analysis at high enough temperatures show that there are no such degrees of freedom.
Also, the assumption that the only phase transition lines are the red and green lines implies that we do not get topological degrees of freedom unless we cross either of these lines. Notice also that we cannot write down mass terms for topological degrees of freedom because they are topological. Therefore, they do not exist also on the red line. 

\section{Conclusion and discussion}
In this paper we have studied an 't~Hooft anomaly in the space of imaginary chemical potential $\mu_B$. 
It is an anomaly between the periodicity $\mu_B \sim \mu_B +2\pi$ of the imaginary chemical potential and the chiral symmetry $\SU(N_f)_L \times \SU(N_f)_R$, and is given by \eqref{eq:anomalyUV}. This anomaly indicates the existence of either at least one phase transition point in the range of $\left[0, 2\pi\right)$ or some gapless degrees of freedom. We have explained how this anomaly is matched by the pions after chiral symmetry breaking. Then we have discussed applications of the anomaly to the QCD phase diagram on the plane $(T,\mu_B)$. 
Possibilities like Figure~\ref{fig:pdimp} are excluded, while possibilities like Figure~\ref{fig:pdpt} and Figure~\ref{fig:pdps} are allowed. 
In particular, we have obtained constraints on what happens on the Roberge-Weiss line, which is the red line in these Figures.

In Figure~\ref{fig:pdps}, the second order phase transition on the red dashed line should be described by some nontrivial (rather than free) CFT if the greatest common divisor $\textrm{gcd}(N_c,N_f)$ of the color number $N_c$ and the flavor number $N_f$ is greater than 1. It is an interesting possibility if it is really realized.
On the other hand, if the situation in Figure~\ref{fig:pdpt} is realized, the phase transition on the red line is first order, at least before it hits the green line on which chiral phase transition happens. There are two possibilities for the intersection point between the red and the green lines. One possibility is that it is some critical point described by a nontrivial CFT.
Another possibility is that the phase transition is first-order also at this point. In the latter case, the phase transition on the green line (chiral phase transition) should also be first-order at least near the intersection point. Then the simplest possibility is that the phase transition is first-order on the entire green line. In particular, if large $N_c$ analysis is valid, the dependence on $\mu_B$ should be small and hence it is natural that the entire green line is first-order.\footnote{As discussed in \cite{Yonekura:2019vyz}, $N_f$ can also be large as far as the ratio $N_f/N_c$ is not too large so that the chiral symmetry is broken at low temperatures.} 
This point is emphasized in \cite{Shimizu:2017asf,Yonekura:2019vyz}.
(See also \cite{Hidaka:2011jj} for another argument based on large $N_c$ and QCD inequalities.) However, constraints from 't~Hooft anomalies are valid even for small $N_c$,
such as $N_c=N_f=3$.

\section*{Acknowledgements}

KY is supported in part by JST FOREST Program (Grant Number JPMJFR2030, Japan), MEXT-JSPS Grant-in-Aid for Transformative Research Areas (A) ``Extreme Universe'' (No. 21H05188), and JSPS KAKENHI (17K14265). SKK is supported by Grant-in-Aid for JSPS Fellows (No. 22KJ0311) from MEXT, Japan. TY is supported by Graduate Program on Physics for the Universe (GP-PU), Tohoku University and JST SPRING, Grant Number JPMJSP2114. 

\renewcommand{\thesection}{\Alph{section}}
\stepcounter{app}

\section{The free energy}\label{sec:freeenergy}
In this appendix we reproduce some of the results in Appendix D of \cite{Gross:1980br}.
The purpose is to derive the free energy formulas \eqref{eq:TotalF}, \eqref{eq:gluonF}, \eqref{eq:quarkF} and \eqref{eq:ferfdef}.

We perform the path integral on the Euclidean spacetime $M_4 = S^1 \times \bR_3$.
Recall that the Lagrangian is 
\begin{align}
\cL= -\frac{1}{4g^2} \tr_c((F_C)_{\mu\nu} (F_C)^{\mu\nu} ) + \overline{\Psi}\gamma^\mu D_\mu \Psi,\
\end{align}
where $(F_C)_{\mu\nu}$ is the field strength of the $\SU(N_c)$ gauge field $(A_C)_\mu$.
The covariant derivative acting on $\Psi$ is given by $D_\mu=\partial_\mu+A_\mu$, where
\beq
	A=  A_C+\frac{1}{N_c}1_{N_c}A_B 
\eeq 
is a $\U(N_c)$ gauge field which combines the dynamical gauge field $A_C$ for $\SU(N_c)$ and the background gauge field $A_B$ for $\U(1)_B$.

We introduce a constant gauge field $\overline{A}_\mu$ which has only the time component,
\begin{align}
	\overline{A}_\mu = \frac{a}{\i \beta}\delta_{\mu0} = a \xi_\mu
\end{align}
where 
\begin{align}
a = \diag(a_1,\cdots,a_{N_c})	\label{eq:diagnala}
\end{align}
 is a diagonal $N\times N$ matrix, and $\xi_\mu := \frac{1}{\i\beta}\delta_{\mu0}$.
Then we separate $A_\mu$ as $A_\mu=\overline{A}_\mu + g B_\mu$ where $B_\mu$ is the fluctuation over which we perform the path integral.
This is the same procedure as in the computation of a Coleman-Weinberg potential or 1PI effective action.

We expand the action to the quadratic order with respect to $\Psi$ and $B_\mu$, and then compute the functional determinant at the one-loop level.
At this order the covariant derivative can be approximated by replacing $A_\mu$ by $\overline{A}_\mu$. More explicitly,
on each component $\Psi_i$ and $(B_\mu)_{ij}$ of the $\SU(N_c)$ vector $\Psi$ and the matrix $B_\mu$, we have
\begin{align}
	(D_\mu\Psi)_i  = (\partial_\mu + a_i\xi_i)\Psi_i, \qquad (D_\mu B_\nu)_{jk} =( \partial_\mu +  (a_j-a_k)\xi_\mu)(B_\nu)_{jk}.
\end{align}

First let us consider the contribution to the free energy from $\Psi$. It is given by
\begin{align}
	F_{\text{quark}}(T,\{a^i\})
		&= -\frac{1}{\beta V_3}\log \Det_-  \slashed{D} =  -\frac{1}{2\beta V_3}\log \Det_- [-D^2] \nonumber \\
		&=-\frac{2N_f}{\beta V_3}\sum_{i=1}^{N_c}\log \Det_- [-(\partial_\mu +  a_i\xi_\mu)^2].	\label{eq:calFquark}
\end{align}
where $\Det_-$ means the determinant in the functional space of $\Psi$ or its components. (In what functional space we consider $\Det_-$ is hopefully clear from the context.)
The subscript $-$ of $\Det_-$ means that the boundary condition on $S^1$ is anti-periodic.
We have taken into account the fact that there are $4 \times N_f$ components from the spinor and flavor indices.

Next we consider the contribution from $B_\mu$. 
We can fix the gauge by introducing a term proportional to $\tr_c (D_\mu B^\mu)^2$. Then we also need to introduce ghosts $c, \overline{c}$. The Lagrangian for $B_\mu, c$ and $\overline{c}$ after the gauge fixing is
\beq
\cL = - \frac{1}{2} \tr_c (D_\mu B_\nu D^\mu B^\nu) + D_\mu \overline{c} D^\mu c.
\eeq
Then, the total contribution is the same as the $ 4 - 2=2$ real scalars (or equivalently one complex scalar) in the adjoint representation, where $4$ comes from $B_\mu$
and $-2$ comes from $(c,\overline{c})$. Thus we get
\begin{align*}
	F_{\text{gluon}} (T,\{a^i\})
		&=  -\frac{1}{\beta V_3} \log \Det_+^{-1}[-D_\mu D^\mu]\\
		&=  \frac{1}{\beta V_3}\left\{(N_c-1)\log \Det_+[-(\partial_\mu)^2]+ \sum_{j\neq k}\log \Det_+[-(\partial_\mu + (a^j-a^k)\xi_\mu)^2]\right\} \label{eq:calFgluon}
		\stepcounter{equation}\tag{\theequation} 
\end{align*}
where the subscript $+$ of $\Det_+$ means that the boundary condition on $S^1$ is periodic.

Thus we have to calculate $\log \Det_{\pm}[-(\partial_\mu+\calq\xi_\mu)^2]$ for $\calq= 0, a^i,a^j-a^k$.
But we only need to calculate the anti-periodic case 
\begin{align}
f(T,\calq):= - \frac{1}{\beta V_3} \log \Det_-[-(\partial_\mu+\calq\xi_\mu)^2] 	\label{eq:antiperiodicPI}
\end{align}
because of the relation
\beq\label{eq:pmrelation}
\log \Det_+[-(\partial_\mu+\calq\xi_\mu)^2] = 	\log \Det_-[-(\partial_\mu+(\calq-\pi) \xi_\mu)^2]
\eeq
which follows from the change of the path integral variable $\Psi(\tau, \vec x) \to \Psi'(\tau, \vec x) = e^{\pi \i \tau/\beta} \Psi (\tau, \vec x)$.
More explicitly, the result for the periodic case is given by
\beq
b(T,\calq):= \frac{1}{\beta V_3} \log \Det_+[-(\partial_\mu+\calq\xi_\mu)^2]  = - f( T, \calq - \pi).
\eeq
By using $f(T,\calq)$ and $b(T,\calq)$, we can express $F_{\text{gluon}} (T,\{a^i\})$ and $F_{\text{quark}}(T,\{a^i\})$ as in \eqref{eq:gluonF} and \eqref{eq:quarkF}.

Our remaining task is to calculate $f(T,\calq)$.
We denote Matsubara frequencies by $\omega_n = 2\pi(n+1/2) \beta^{-1}~(n \in \bZ ) $. Then
\begin{align*}
	f(T,\calq)
		= - \frac{1}{\beta V_3} \Tr \log [-(\partial_\mu+\calq\xi_\mu)^2] 
		= -\frac{1}{\beta} \sum_{n \in \bZ} \int \frac{d^3 k}{(2\pi)^3}   \log\left[\left(\omega_n-\frac{\calq}{\beta}\right)^2+\vec{k}^2\right].
		\stepcounter{equation}\tag{\theequation} 
\end{align*}
By using the Poisson resummation formula~\footnote{This formula is the Fourier mode expansion of the periodic function $  \sum_{n \in \bZ} \delta(x-n) $ in terms of
the Fourier modes $e^{2\pi \i m x}$. One can check that the Fourier coefficients are all 1.}
\beq
  \sum_{n \in \bZ} \delta(x-n) =   \sum_{m \in \bZ} e^{2\pi \i m x},
\eeq
where $\delta(x)$ is the delta function, we obtain
\beq
f(T,\calq)= -\sum_{m \in \bZ}(-1)^m \int \frac{d^4 k}{(2\pi)^4} e^{\i m \beta k^0}  \log\left[\left(k^0-\frac{\calq}{\beta}\right)^2+\vec{k}^2\right].
\eeq
The term with $m=0$ does not depend on $T$ nor $\calq$, so it is just the contribution to the 4d cosmological constant term and hence we neglect it. 
Then, by (i) shifting the integration variable $k^0 \to k^0+\calq/\beta$, (ii) doing integration by parts with respect to $k^0$, and (iii) performing the integral over $k^0$ by closing the integration contour depending on the sign of $m$, we get
\beq
f(T,\calq) &= -\sum_{m \neq 0}(-1)^m e^{\i m \calq} \int \frac{d^4 k}{(2\pi)^4} e^{\i m \beta k^0}  \log\left[(k^0)^2+\vec{k}^2\right] \nonumber \\
&= \sum_{m \neq 0}(-1)^m e^{\i m \calq} \int \frac{d^4 k}{(2\pi)^4} \frac{e^{\i m \beta k^0}}{\i m \beta } \frac{2k^0}{(k^0)^2+\vec{k}^2} \nonumber \\
&= \sum_{m =1}^\infty (-1)^m (e^{\i m \calq}+e^{-\i m \calq}) \int \frac{d^3 k}{(2\pi)^3} \frac{e^{- m \beta |\vec k|}}{ m \beta } \nonumber \\
&= \frac{2}{\pi^2} T^4 \sum_{m =1}^{\infty} (-1)^m  \frac{\cos m\calq }{m^4}.
\eeq
In terms of the polylogarithm 
	\begin{align}
		\text{Li}_s(z):=\sum_{n=1}^\infty\frac{z^n}{n^{s}},
	\end{align}
we can write the result as
\beq
f(T,\calq)  = \frac{2}{\pi^2} T^4 \sum_{n =1}^{\infty} (-1)^n  \frac{\cos n\calq }{n^4}= \frac{2}{\pi^2} T^4 \Re \text{Li}_s(-e^{\i \calq}).
\eeq	

It is known that $\text{Li}_s(z)$ has a branch point at $z=1$, and hence $f(T,\calq)$ is not analytic at $\calq \equiv \pi \mod 2\pi $. More explicitly we have
	\begin{align}
		\frac{\partial^{s-1}}{\partial\calq^{s-1}}\text{Li}_s(-e^{\mathsf{i}\calq})=\i^{s-1}\sum_{n=1}^\infty\frac{(-e^{\mathsf{i}\calq})^{n}}{n}
		=-\i^{s-1}\text{Log}\left(1+e^{\i \calq }\right)
	\end{align}
where $\text{Log}$ is the principal value of the logarithm. The real part of this is indeed discontinuous at  $\calq \equiv \pi \mod 2\pi $.
Because $\text{Li}_4\left(-e^{\mathsf{i}\calq}\right)$ is an analytic function in the range $-\pi<\calq<\pi$, the function $f(T,\calq)$ can be Tayler-expanded around $\calq=0$. 
Notice that
	\begin{align}
		\frac{\partial^{4}}{\partial\calq^{4}} \Re \text{Li}_4(-e^{\mathsf{i}\calq})
		=- \Re \frac{e^{\i \calq} }{e^{\i \calq}+1} = -\frac{1}{2} ,
	\end{align}
and in particular it is independent of $\calq$. Thus the derivatives $\frac{\partial^\ell}{\partial \calq^\ell}f|_{\calq =0}$ for $\ell>4$ all vanish.
We can calculate the derivatives for $\ell  < 4$ by using the expression $\sum_{n =1}^{\infty} (-1)^n  \frac{\cos n\calq }{n^4}$ and find
\beq
f(T,\calq) =  \frac{2}{\pi^2} T^4 \left[ -\frac{7}{8}\zeta(4) + \frac{1}{2!} \cdot \frac{1}{2}\zeta(2) \calq^2 +  \frac{1}{4!} \cdot \left( -\frac{1}{2} \right) \calq^4  \right] ,
\eeq
which gives the result \eqref{eq:ferfdef}.

\section{The minimum of the free energy}\label{sec:min}
In this appendix we show that the minimum of the free energy (\ref{eq:TotalF}) under the constraint (\ref{eq:aiconstraint}) is given by \eqref{eq:a=a} where $n$ is within the range 
$-\pi \leq \mu_B+2\pi n \leq \pi$.

First let us observe the following point.
Suppose that the quark free energy $F_{\text{quark}}(T,\{a_i\})$ has a minimum at 
\beq\label{eq:minA}
	a_1 = \cdots = a_{N_c} = \frac{\mu_B+2\pi n}{N_c} \mod 2\pi \bZ 
\eeq
for some $n$. Then we can immediately conclude that this point is also a minimum point of the total free energy
$F(T,\{a_i\}) = F_{\text{gluon}}(T,\{a_i\})+F_{\text{quark}}(T,\{a_i\})$, because \eqref{eq:minA} also gives one of the absolute minimum points of the gluon free energy 
$F_{\text{gluon}}(T,\{a_i\}) $ even if we forget the constraint \eqref{eq:aiconstraint},
as one can check by using the formula \eqref{eq:gluonF}. Therefore, we study $F_{\text{quark}}(T,\{a_i\})$ in the following. 

To make expressions slightly simpler, we will use
\beq
\tilde \mu_B := \frac{\mu_B}{\pi}, \quad \tilde a_i := \frac{a_i}{\pi},
\eeq
$\mu_B$ and $a_i$ are $2\pi$-periodic variables, and hence they can be restricted to the range
\beq
-1 \leq \tilde \mu_B \leq 1, \quad  -1 \leq \tilde a_i \leq 1.
\eeq
We also define
\beq
  \tilde f (\tilde \mu) :=  \frac{24}{\pi^2 T^4}f(T,  \mu)    = - \frac{7}{15}+  2 \tilde \mu^2 - \tilde \mu^4 \qquad ( -1 \leq \tilde \mu \leq 1).
\eeq
Then $F_{\text{quark}}(T,\{a_i\})$ given in \eqref{eq:quarkF} is rewritten as
\begin{align}
	F_{\text{quark}}(T,\{a_i\})= \frac{\pi^2N_f  T^4}{12}  \sum_{i=1}^{N_c} \tilde f(\tilde a_i) .
	\end{align}
 The constraint \eqref{eq:aiconstraint} is 
 \beq\label{eq:constB}
 \sum_{i=1}^{N_c} \tilde a_i = \tilde  \mu_B +2n, \qquad  n \in \bZ.
 \eeq

Observe that $\tilde f(\tilde \mu)$ is a monotonically increasing function of the absolute value of $\tilde \mu$, within the range $|\tilde \mu| \in [0,1]$.
Now we show several properties of minimum points of $ \tilde F_{\text{quark}}(\{\tilde a_i\}) $.
\begin{itemize}

\item 
A minimum is realized when the $n$ in \eqref{eq:constB} satisfies $| \tilde{\mu}_B +2n | \leq 1$. It can be shown as follows. Assume on the contrary that 
$| \tilde \mu_B +2n |>1$. Then we define $C := \tilde \mu_B/(\tilde \mu_B +2n) $. Given a point $\{\tilde a_i\}$ satisfying $ \sum_{i=1}^{N_c} \tilde a_i = \tilde  \mu_B +2n$, we have a different point $\{C \tilde a_i\}$ satisfying $ \sum_{i=1}^{N_c}  C \tilde a_i = \tilde  \mu_B $. This new point has a lower value of $  F_{\text{quark}} $ than the one at the original point due to the monotonicity of the function $\tilde f(\tilde \mu)$ and the fact that $|C|<1$ (recall $|\tilde \mu_B| \leq 1$).

\item 
At a minimum point, all nonzero $\tilde a_i$ have the same sign. It can be shown as follows. Assume on the contrary that $\tilde a_j>0$ and $\tilde a_k<0$ for some $j,k$. 
Let $\epsilon$ be a small positive number. Then, we take a new point $\{\tilde a'_i\}$ with $\tilde a'_j = \tilde a_j -\epsilon$, $\tilde a'_k = \tilde a_k+\epsilon$, and 
$\tilde a'_i=\tilde a_i $ for $i \neq j,k$.
The new point satisfies the same constraint as the original one, $\sum_i \tilde a'_i = \sum_i \tilde a_i$. 
By taking $\epsilon$ sufficiently small, we have $|\tilde a'_j |< |\tilde a_j |$ and $|\tilde a'_k| < |\tilde a_k | $. By the monotonicity of $\tilde f(\tilde \mu)$, the new point has a lower value of $  F_{\text{quark}} $ than the one at the original point.

\item
At a minimum point, all $\tilde a_i$ are the same. It can be shown as follows. Assume on the contrary that $\tilde a_j \neq \tilde a_k$ for some $j,k$.
Let $t$ be a variable within the range $t \in [0,1]$. Then, we take a new point $\{\tilde a'_i\}$ with 
\beq
\tilde a'_j =\tilde a'_j(t):= (1-t) \tilde a_j +t \tilde a_k, \qquad \tilde a'_k = \tilde a'_k(t) :=(1-t) \tilde a_k + t\tilde a_j, 
\eeq
and 
$\tilde a'_i=\tilde a_i $ for $i \neq j,k$.
The new point satisfies the same constraint as the original one, $\sum_i \tilde a'_i = \sum_i \tilde a_i$. 
Now we study a minimum of the function
\beq
g(t): = \tilde f( \tilde a'_j(t)) + \tilde f(\tilde a'_k(t)).
\eeq
Its derivative with respect to $t$ is
\beq
\frac{dg}{dt} = -4(1-2t)(\tilde a_j - \tilde a_k)^2 \left[ 1 - ( \tilde a'^2_j + \tilde a'_j \tilde a'_k + \tilde a'^2_k)  \right].
\eeq
We have shown above that all nonzero $\tilde a_i$ have the same sign. We have also shown that a minimum is achieved only if $|\tilde \mu_B+2n| \leq 1$.
These two facts imply that 
\beq
 \tilde a'^2_j + \tilde a'_j \tilde a'_k + \tilde a'^2_k \leq  (\tilde a'_j +  \tilde a'_k)^2 \leq \Big( \sum_i \tilde a'_i \Big)^2 \leq 1.
\eeq
The equality holds only if either $ |\tilde a'_j |$ or $ |\tilde a'_k|$ is 1 and all others are zero.
Therefore, $g(t)$ has the minimum at $t=1/2$ with a lower value of $  F_{\text{quark}} $ than the one at the original point.

\end{itemize}
By the above results, we conclude that $\{\tilde a_i\}$ at a minimum point is given by
\beq
\tilde a_1 = \cdots = \tilde a_{N_c} = \frac{\tilde \mu_B+2 n}{N_c}, \qquad  -1 \leq \tilde  \mu_B+2 n\leq 1.
\eeq
For generic values of $\tilde \mu_B$, there is only a single value of $n$ in the above range. 
However, when $\tilde \mu_B \equiv 1 \mod 2$, (i.e., $ \mu_B \equiv \pi \mod 2\pi$), there are two minima of $  F_{\text{quark}} $.

\section{Impossibility of free 3d fermions}\label{sec:technicalproof}
In this appendix, we prove \eqref{eq:multn} which implies that the second order phase transition at the red dashed line in Figure~\ref{fig:pdps} is not described by free 3d fermions.

Let $T$ be the element of the Lie algebra of $\SU(N_f)$ given by\footnote{Unlike the rest of the paper, we take $T$ to be hermitian rather than anti-hermitian.}
\begin{align}
	T = \frac{1}{N_f}
		\left(
			\begin{array}{cccc}
				1&&&\\
				&\ddots&&\\
				&&1&\\
				&&&1-N_f
			\end{array}
		\right).
\end{align}
Then the element of the $\SU(N_f)$ 
\begin{align}
	e^{2\pi \i T} = e^{2\pi \i/N_f}I_{N_f}
\end{align}
is a generator of the center of $\SU(N_f)$.
In irreducible representations $\rho_{L}$ and $\rho_{R}$,
this center element should be proportional to the unit matrix by Schur's lemma, and hence there are integers $n_L, n_R$ such that
\begin{align}
	\rho_{L,R}\left(e^{2\pi \i T} \right) = e^{2\pi \i n_{L,R}/N_f}I_{d_{L,R}}.	\label{eq:rhoeT}
\end{align}
Then, $\rho_L(T)$ (after using a unitary transformation to make it diagonal) should be of the form,
\begin{align}
	\rho_L(T) = \left(
			\begin{array}{cccc}
				\frac{n_L}{N_f} + \ell_1&&&\\
				&\frac{n_L}{N_f} + \ell_2&&\\
				&&\ddots&\\
				&&&\frac{n_L}{N_f} + \ell_{d_L}
			\end{array}
		\right),
\end{align}
where $\ell_i \in \bZ \ (i=1,\cdots,d_L)$. 
Since $\det \rho_L$ is a trivial representation of $\SU(N_f)$,\footnote{It is a one-dimensional representation of $\SU(N_f)$, and the only one-dimensional representation of $\SU(N_f)$ is the trivial representation.} we have
\begin{align}
	0 = \tr \rho_L(T)=\frac{n_L}{N_f} d_L + \sum_{i=1}^{d_L} \ell_i.	\label{eq:detrhoL}
\end{align}
We get a similar result about $d_R$,
\begin{align}
	\frac{n_R}{N_f}d_R + \sum_{i=1}^{d_R} r_i = 0	\label{eq:detrhoR},
\end{align}
where $r_i\in \bZ$.

By using the above results, we compute $d_R k_L$. The Dynkin index $k_L$ is
\begin{align}
	k_L = \frac{\tr_{\rho_L}T^2}{\tr_{\rho_f}T^2} 
		= \frac{N_f}{N_f-1}\sum_{i=1}^{d_L} \left(\frac{n_L}{N_f} + \ell_{i}\right)^2
		= \frac{N_f}{N_f-1}\left(\frac{n_L}{N_f}\sum_{i=1}^{d_L}\ell_i + \sum_{i=1}^{d_L}\ell_i^2 \right).
\end{align}
Then
\begin{align*}
	d_R k_L
		&= \frac{1}{N_f-1}\left(d_Rn_L\sum_{i=1}^{d_L}\ell_i + d_RN_f\sum_{i=1}^{d_L}\ell_i^2 \right)\\
		&= \frac{1}{N_f-1}\left(d_R(n_L+n_R)\sum_{i=1}^{d_L}\ell_i -d_Rn_R\sum_{i=1}^{d_L}\ell_i + d_RN_f\sum_{i=1}^{d_L}\ell_i^2 \right) \label{eq:dRkL}
		\stepcounter{equation}\tag{\theequation} 
\end{align*}
From (\ref{eq:detrhoR}), we see that $d_Rn_R$ is a multiple of $N_f$. Also recall that the representation $\rho_L\otimes \rho_R$ is invariant under the $\bZ_n$ which appears in \eqref{eq:Zn} and hence we should have
\begin{align}
	 \rho_L(e^{2\pi \i/n }I_N) \otimes \rho_R(e^{2\pi \i/n }I_N) = e^{2\pi \i\frac{n_L+n_R}{n}} I_{d_{\rho_L}}\otimes I_{d_{\rho_R}} = I_{d_{\rho_L}}\otimes I_{d_{\rho_R}}.
\end{align}
Then, $n_L+n_R$ is a multiple of $n$. Finally, $N_f$ is a multiple of $ n$ by the definition $n=\text{gcd}(N_c,N_f)$.
Therefore each term inside the bracket of (\ref{eq:dRkL}) is a multiple of $n$. 
Because $n$ and $N-1$ are coprime, we conclude that $d_Rk_L$ is a multiple of $n$.

\bibliographystyle{ytphys}
%\baselineskip=.95\baselineskip
\bibliography{ref}

\end{document}